\documentclass[twocolumn,showpacs,preprintnumbers,amsmath,amssymb,10pt,a4paper]{revtex4}

\usepackage{geometry}
\usepackage{graphicx}
\usepackage{dcolumn}
\usepackage{bm}
\newcommand{\be}{\begin{equation}}
\newcommand{\ee}{\end{equation}}
\newcommand{\bd}{\begin{displaymath}}
\newcommand{\ed}{\end{displaymath}}
\newcommand{\BE}{\begin{eqnarray}}
\newcommand{\EE}{\end{eqnarray}}

\newcommand{\avg}[1]{\left\langle{#1}\right\rangle}

\newcommand{\Da}{\mbox{Da}}
\begin{document}

\preprint{}
\title{Time scales and species coexistence in chaotic flows}

\author{Tobias Galla}
\email{tobias.galla@manchester.ac.uk}
\affiliation{Theoretical Physics, School of Physics and Astronomy, The University of Manchester, Manchester M13 9PL, United Kingdom}

\author{Vicente P\'erez-Mun\~uzuri}
\email{vicente.perez@cesga.es}
\affiliation{Group of Nonlinear Physics, Faculty of Physics, University of Santiago de Compostela, E-15782 Santiago de Compostela, Spain}

\date{\today}

\begin{abstract}
Empirical observations in marine ecosystems have suggested a balance of biological and advection time scales as a possible explanation of species coexistence. To characterise this scenario, we measure the time to fixation in neutrally evolving populations in chaotic flows. Contrary to intuition the variation of time scales does not interpolate straightforwardly between the no-flow and well-mixed limits; instead we find that fixation is the slowest at intermediate Damk\"ohler numbers, indicating long-lasting coexistence of species. Our analysis shows that this slowdown is due to spatial organisation on an increasingly modularised network. We also find that diffusion can either slow down or speed up fixation, depending on the relative time scales of flow and evolution.
\end{abstract}


\maketitle
{\em Introduction.}
How biological systems maintain species coexistence is still far from understood. Gause's exclusion principle suggests that two species competing for the same resources cannot coexist in the long term \cite{gause,hardin}. Explanations of the biodiversity observed in real-world ecosystems often involve a combination of spatial separation and local interaction. Niches form, hosting different species at distinct locations. Species coexistence is however also found in stirred and seemingly homogeneous oceanic systems \cite{paradox}. Horizontal mixing and turbulent flows have been offered as possible explanations, supported by theoretical models \cite{karolyi2000, young,neufeld2,neufeld,groselj} and satellite observations of chlorophyll concentrations off the coast of South America \cite{dovidio,kudela}. Broadly speaking a flowing medium generates stable niches, spatially separated, and sufficiently long lived to let individual organisms thrive. 

This naturally raises the question of time scales. The flow must be strong enough for the heterogeneity to be biologically noticeable. On the other hand stirring must not be too quick, as premature mixing would impede coexistence. Several studies suggest that patchiness in plankton populations is related to the Damk\"ohler number characterising the ratio between the time scales associated with the flow and the biological processes respectively \cite{pasquero, neufeld2, young}.

Theoretical models have been used to study the dynamics of species coexistence in flowing environments \cite{neufeld2, neufeld, groselj, pigolotti2012,benzi2012,perlekar2013,pigolotti2013,pigolotti2014}. Some of this work focuses on the emerging spatial structures, including the formation of filaments in chaotic flows \cite{abraham, karolyi2000,young}. Only a relatively small number of studies have systematically investigated the role of the time scales governing advection and evolution, see e.g. \cite{young,neufeld,groselj}. The aim of our work is to address this gap, and to use a stylised model to determine how long coexistence is maintained at different Damk\"ohler numbers. 

Each individual in our model belongs to one of two species; these particles are advected by a chaotic flow. Our analysis focuses on the mean time until one species is eliminated, known as the time to fixation in genetics and evolutionary biology \cite{ewens, nowak}. In each evolutionary step one particle is selected for removal, and a randomly chosen particle in its vicinity reproduces, similar to dynamics of the well-known voter model (VM) \cite{liggett}. Neutral models of this type have been used in ecology, but also in the context of opinion dynamics and language evolution, for examples see \cite{fortunato, blythe1}. The combination of a finite interaction range and the underlying flow leads to a constantly changing interaction network. Adaptive networks in existing studies of VMs typically evolve through a process of rewiring, driven by the individuals at its nodes \cite{gross, fragmentation}. In contrast, the network in our model is shaped by the flow. The positions of the particles are not constrained to a discrete lattice, and a heterogeneous network can form, distinguishing our work from that of \cite{karolyi2000}. 

Varying the Damk\"ohler number interpolates between two extremes: a static network in the limit of infinitely fast evolution, and an effectively well-mixed system for very fast flow. Fixation times in simple well-mixed systems can be obtained analytically with tools from statistical physics, and it is well known that fixation in the VM happens much sooner in a well-mixed situation than in static networks \cite{sood, bennaim}. 

Surprisingly, our analysis shows that the interpolation between these regimes is not necessarily monotonic; fixation can be slowest when the time scales of flow and evolution balance. This indicates that moderate stirring can maintain species coexistence for longer than complete mixing or the absence of any flow.

{\em Model.} The model describes a population of $N$ individuals in a two-dimensional domain $[0,1]^2$ with periodic boundary conditions. The size of the population is constant. We write $x_i(t)$ and $y_i(t)$ for the coordinates of particle $i$. At each time, each particle can be in one of two states, $\sigma_i(t)\in\{0,1\}$, representing the species the individual belongs to. The key components of the model are (i) a flow field advecting the particles, and (ii) the evolutionary birth-death process. 
The aim of our work is not to study the detailed effects of different chaotic flows. Instead we choose a periodic parallel shear flow as a simple representative, displaying repeated stretching and folding and mimicking turbulent flow \cite{shear1,shear2,shear3}. The individuals are treated as Lagrangian particles; we write $\dot x_i=s_1(x_i,y_i,t), ~ \dot y_i=s_2(x_i,y_i,t)$, where the quantities $\dot x_i$ and $\dot y_i$ are derivatives with respect to $t$. During the first half of each period the shear flow is given by
\be
s_1(x_i,y_i,t)=V_0\sin[2\pi y_i+\phi],~s_2(x_i,y_i,t)= 0.
\ee
We choose a period of one, $V_0$ sets the flow velocity. During the second half of each period one has
\be
s_1(x_i,y_i,t)=0, ~ s_2(x_i,y_i,t)=V_0\sin[2\pi x_i+\phi].
\ee
The phase $\phi$ is drawn at random from a uniform distribution over $[0,2\pi)$ for each period. These equations of motion are integrated by standard methods. We compare the outcome for this structured flow with that of entirely random diffusion
\be\label{eq:noise}
\dot x_i =\xi_i, \dot y_i=\eta_i,
\ee
where the $\xi_i, \eta_i$ are independent Gaussian processes of mean zero. We also consider interpolations between structured flow and random diffusion,
\be\label{eq:flow}
\dot x_i=\sqrt{1-\alpha^2}s_1+\alpha \xi_i, ~ 0\leq \alpha \leq 1,
\ee
and similar for $\dot y_i$. The amplitude of the noise flow is chosen such that the root mean square velocity is constant for all $\alpha$, i.e., $\avg{\dot x_i^2+\dot y_i^2}^{1/2}=V_0/2$, with $\avg{\cdots}$ a spatial average.
\begin{figure}[t!!!]
\begin{center}
\vspace{-1cm}
  \hspace{-0.5cm}\includegraphics[scale = 0.4]{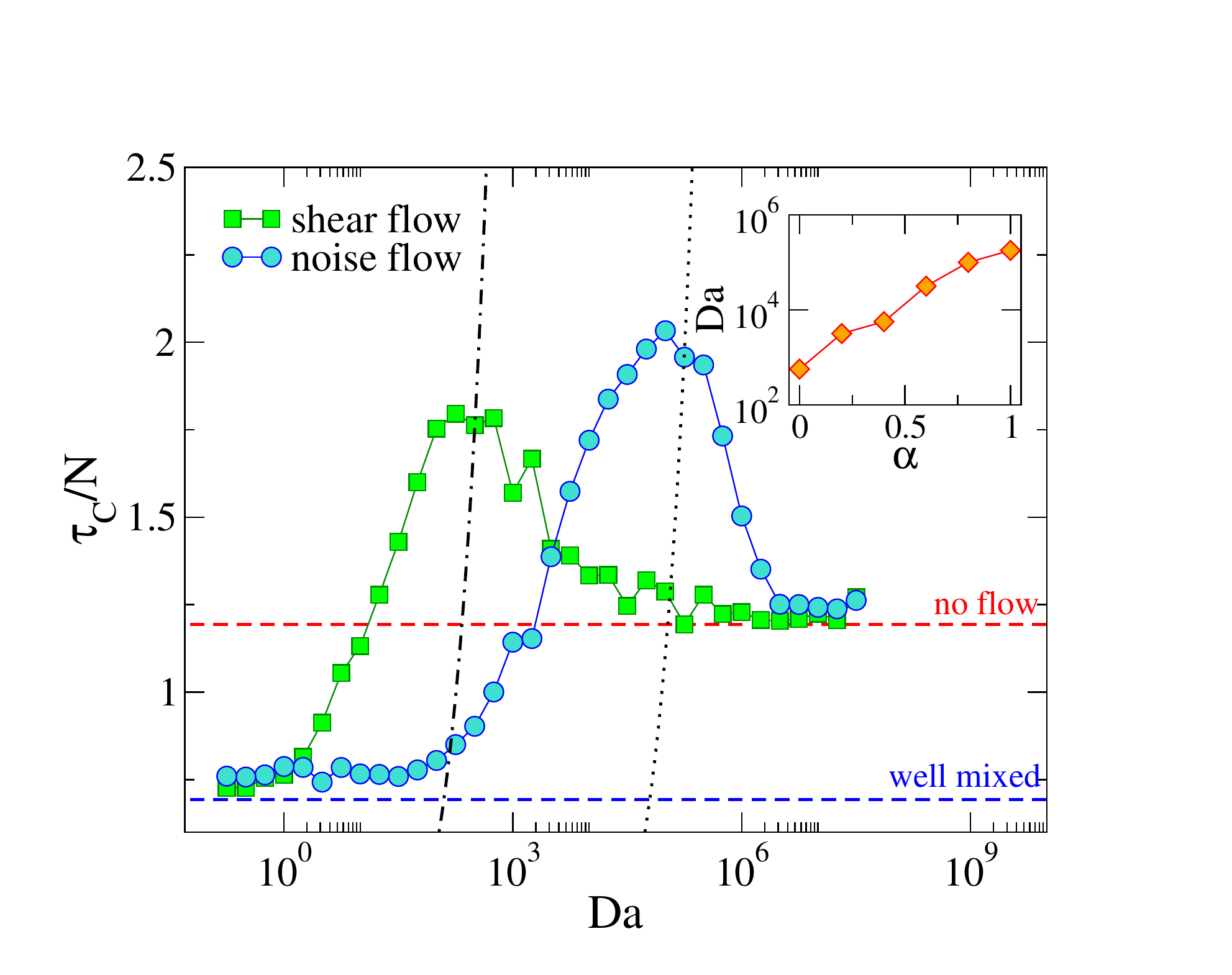}
  \vspace{-1cm}
  \end{center}
  \caption{(Color on-line). Fixation time as a function of Damk\"ohler number (average over $1000$ runs). The initial network is a square grid ($n=30$), particles types are assigned at random initially with equal probability. Dot-dashed line is $\tau=5\Da$, dotted line $\tau=0.01 \Da$, see text. Model parameters are $R=1.5, V_0=1.4$. Inset: Damk\"ohler number for which fixation takes the longest, for different mixtures of structured flow and noise flow [Eq.~(\ref{eq:flow})].}
  \label{fig:fig1}
\end{figure}

Interaction between particles occurs within a radius $\delta$. For given particle positions this defines an interaction network. Particles reside at the nodes of the graph, and two particles are connected by an edge whenever their Euclidean distance is at most $\delta$. Writing the population size as $N=n^2$ is particularly convenient when the initial positions form a regular $n\times n$ lattice, embedded in the spatial domain $[0,1]^2$. It is then useful to write $\delta=R/n$; $R>0$ measures the interaction radius in units of lattice spacings. The birth-death dynamics begins at the same time as the flow, and proceeds as follows: in each microscopic step we select one particle, $i$, at random for removal. Of all particles within radius $\delta$ around $i$ one particle $j$ is selected at random for reproduction. This reflects neutral selection. The offspring is of the same type as its parent, and is placed at position $(x_i,y_i)$. Effectively particle $i$ has adopted state $\sigma_j(t)$. This corresponds to the well known voter model \cite{liggett}, albeit on a dynamic network shaped by the flow.

The relative time scales between flow and evolution are controlled by the Damk\"ohler number $\Da$. We execute one sweep of $N$ microscopic evolutionary steps every $1/\Da$ units of time, $t$. The quantity $\Da$ is the average number of reproduction events each particle undergoes per unit time. We write $\tau=\Da\,t $ for the number of evolutionary sweeps by $t$.

 \begin{figure*}[t!!!]
\hspace{-0.5cm}\includegraphics[scale=0.425]{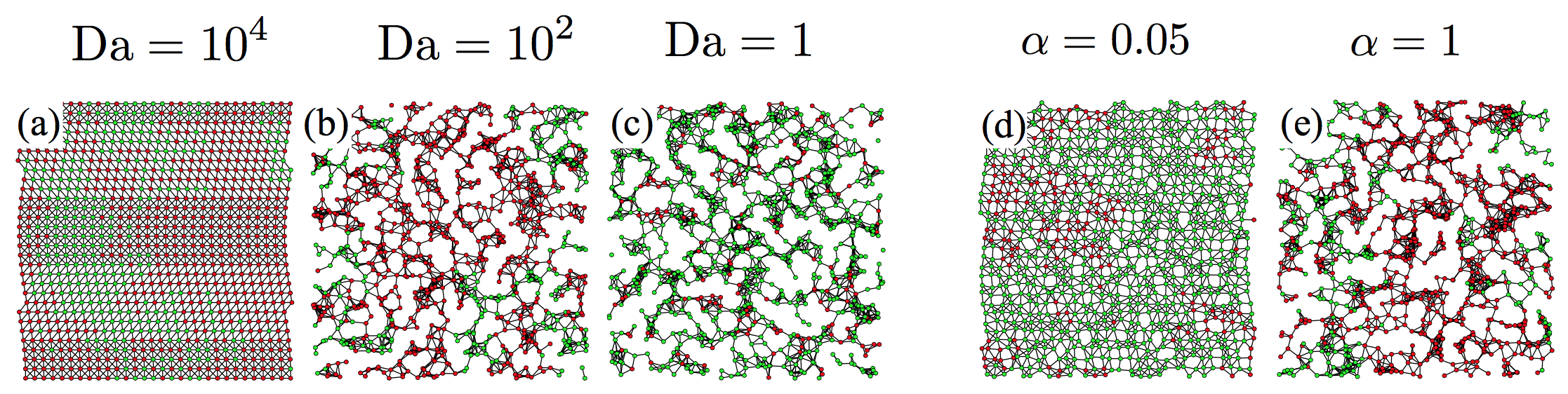}
  \caption{(Color on-line). Examples of the interaction network. Node color (shading) represents the two species. Panels (a)-(c) show snapshots for the structured flow ($\alpha=0$);  flows in (d) and (e) contain noise as indicated; $\Da=10^4$ in (d) and (e). The order parameter $\rho$ takes values $0.34, 0.50, 0.80, 0.59, 0.41$ in panels (a)-(e) respectively. Parameters and initial conditions as in Fig.~\ref{fig:fig1}; snapshots taken at $\tau=200$.}
  \label{fig:fig2}
\end{figure*}

Except for static networks with several disconnected components, the evolutionary dynamics will eventually come to an end when all particles are of the same type. This process occurs through the formation of domains of particles of the same type. These domains gradually coarsen until one domain extends across the entire system; the surviving species has reached fixation. Our main interest is the time it takes to reach this state.

{\em Results.} Our principal result is shown in the main panel of Fig.~\ref{fig:fig1}. Starting from a regular grid we have measured the time to fixation, $\tau_C$, as a function of the Damk\"ohler number. Consistent with the intuitive expectation the limits of small and large Damk\"ohler numbers reproduce results for the well-mixed and no-flow cases respectively ($\tau_C/N=\ln 2$ for the well-mixed system \cite{sood}; $\tau_C/N\approx 1.19$ from simulations on a static $30\times 30$ lattice, $R=1.5$). Intriguingly, however the interpolation between these limits is not monotonic; we observe a maximum in time to fixation at intermediate Damk\"ohler number. This is found for the shear and the noise flow; the location of this maximum varies with the interpolation parameter $\alpha$, see the inset of Fig.~\ref{fig:fig1}. It is these observations that we seek to explain in the following.

{\em Interpretation.}  The data for the shear flow in Fig.~\ref{fig:fig1} shows that the time to fixation is that of the no-flow case for Damk\"ohler numbers above $\sim 10^5$. The following order-of-magnitude calculation helps to develop physical insight. The network changes independently of the evolutionary dynamics. For the parameters in Fig.~\ref{fig:fig1} first significant departures from the regular grid are seen at $t_d\approx 0.01$ \cite{remark1}. Moreover, $\tau_C/N\approx 1$ for all Damk\"ohler numbers. Using this and $N\approx 10^3$ leads to $t_C= t_d$ at $\Da\approx 10^5$.  We conclude that fixation is reached for $\Da\gtrsim 10^5$ before the shear flow has noticeably distorted the initial grid; the fixation time is that of the VM on a static two-dimensional lattice ($R=1.5$) as indicated in Fig.~\ref{fig:fig1}.

The network evolves indefinitely, but will eventually reach a dynamic stationary state. For the shear flow we have measured the half life of edges as $t_\ell\approx 0.5-0.9$. For $N=900$ particles the graph typically contains a few thousand links. Applying again a rough argument, we estimate that the majority of links will have undergone changes after about ten half lives. On those time scales we expect a renewal of the entire network, broadly consistent with the observation that link density and clustering coefficient reach their stationary values at $t^\star\approx 5$ \cite{remark1}. The dot-dashed line in Fig. \ref{fig:fig1} indicates this time scale in evolutionary units, $\tau^\star=\Da\, t^\star$. As seen in the figure $\tau^\star\approx \tau_C$ at the point of maximal slowdown. Using again $\tau_C/N\approx 1$ and $t^\star=10$ for an order-of-magnitude estimate we find $\Da\approx 10^2$ as the point at which evolution and network renewal occur on the same time scale, consistent with the location of the maximum.

The following physical picture emerges. At very slow flow, $\Da\gtrsim10^{5}$, the evolutionary process effectively happens on the network set by the initial condition. At Damk\"ohler numbers just below $\Da=10^{5}$ the coarsening dynamics occurs on a slowly changing network, with only minor distortions from the initial grid, see Fig.~\ref{fig:fig2}(a) for an example. For $10^{2}\lesssim \Da\lesssim 10^{5}$ the birth-death dynamics proceeds adiabatically on a changing network, advected by the flow from the initial grid towards more disorder. This leads to a slowdown of fixation. As a control experiment we have simulated the VM on the largest connected component of {\em static} networks obtained by applying random displacements of the order of the lattice spacing to the nodes of the grid. This disorder increases the time to fixation. The reason for this slowdown is a spatial organization on an increasingly modularised network. For the advected case this can be seen in Fig.~\ref{fig:fig2}(b); patches of particles of the same type are found, broadly corresponding to highly-connected modules of the network. Coarsening occurs at the interfaces of domains, and the relatively low connectivity between patches hinders fixation.

\begin{figure}[t!!!]
\centering
\hspace{-4.7em}  \includegraphics[scale = 0.36]{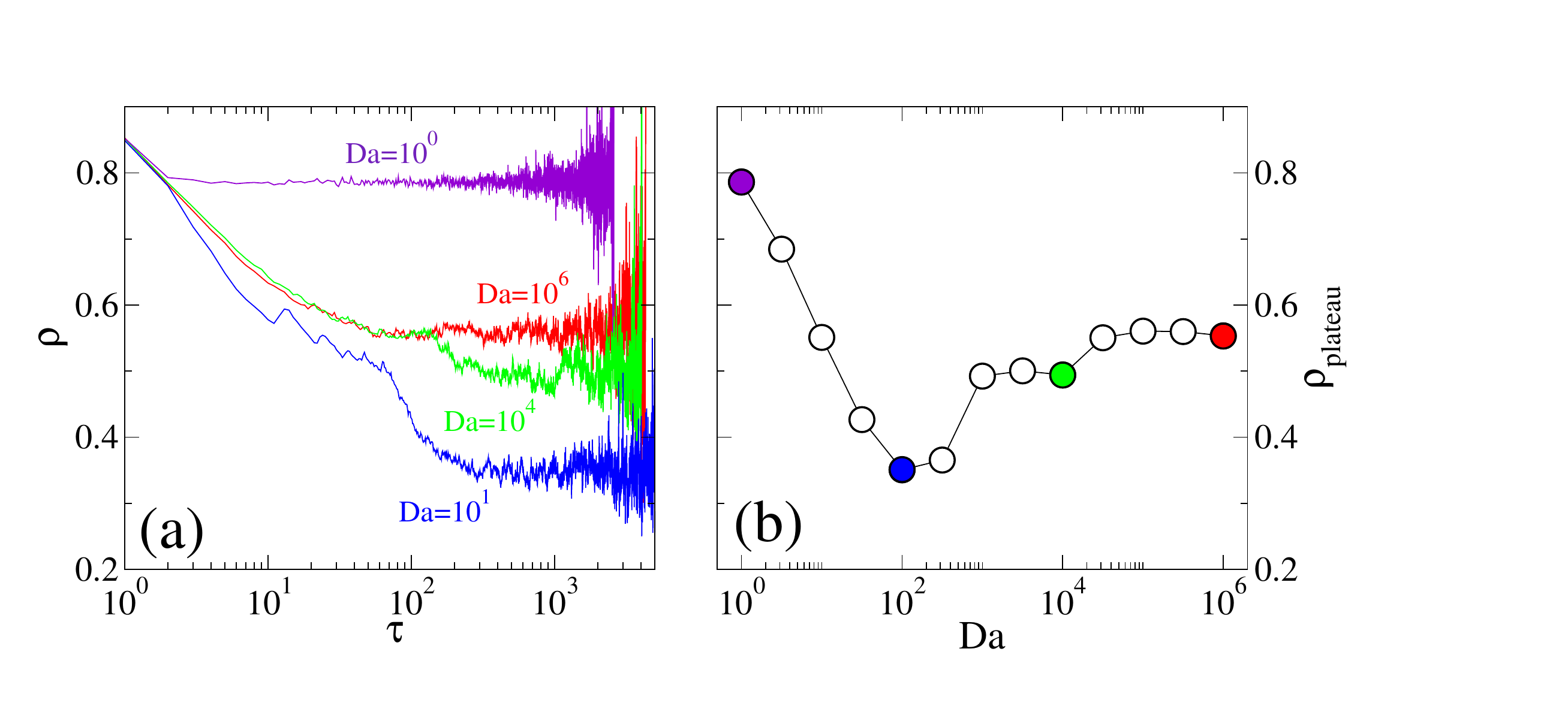}~~~
 \vspace{-0.5cm}
  \caption{(Color on-line). (a) Order parameter $\rho$ [Eq.~(\ref{eq:rho})] as a function of time, $\tau$, measured in sweeps. Data are averages over samples that have not yet reached fixation. (b) Plateau value of $\rho$ for different Damk\"ohler numbers, obtained as a time average of $\rho$ in the interval $200<\tau\leq 500$. Model parameters and initial conditions as in Fig.~\ref{fig:fig1}. }
  \label{fig:fig3}
\end{figure}

To quantify this spatial organisation we measure the fraction $m$ of type-$1$ particles in the system and the probability that a randomly chosen link connects two particles of opposite types, i.e., the fraction of active links, $\rho_a$. We then compute
\be\label{eq:rho}
\rho=\frac{\rho_a}{2m(1-m)}.
\ee
The denominator is known as the \emph{heterozygosity} in genetics \cite{pigolotti2013}, and represents the probability that two individuals chosen at random without regard of the network are of opposite types. Fig.~\ref{fig:fig3}(a) shows that the order parameter $\rho$ initially decreases in the birth-death processes, as spatial domains form. It reaches a plateau before fixation; the numerical value of $\rho$ at the plateau is an indicator of the spatial organisation in this quasi-stationary state. In absence of any spatial correlation one would find $\rho=1$; smaller values indicate clustering of particles of the same type on neighbouring nodes in the network, as shown in Fig.~\ref{fig:fig2}(b). As seen in Fig.~\ref{fig:fig3}(b) maximal spatial organisation occurs for intermediate Damk\"ohler numbers, $\Da\approx 10^{2}$. This clustering slows down the coarsening process, consistent with the maximum observed for the time to fixation in Fig.~\ref{fig:fig1}.

We next discuss the regime of fast flow. For $\Da\lesssim 10^{2}$, the network is no longer quasi-static on the time scale of the birth-death dynamics. The graph is still disordered at any one time, but the evolutionary dynamics cannot organise on the rapidly changing graph. This is seen in Fig.~\ref{fig:fig2}(c), where the extinction process has advanced, but where the minority species is scattered across the network. Hence the plateau value of the order parameter $\rho$ is increased (Fig.~\ref{fig:fig3}). For very fast stirring  the network changes so quickly that particles effectively interact with partners newly sampled at random from the entire population at each evolutionary event. We recover the consensus time of the well mixed VM, see Fig.~\ref{fig:fig1}.

{\em Noise flow.} For pure noise flow the survival time of a link is $t_\ell\approx10^{-3}$; equilibration of the network occurs at $t^\star\approx 0.01$. This indicates a speed-up of network turnover compared to the structured flow, as intuitively expected. The shear flow is relatively coherent and parallel to the axes. In the noise flow particle movements are uncorrelated, implying faster mixing.  Comparing panels (d) and (e) in Fig.~\ref{fig:fig2} against panel (a) demonstrates this; these three snapshots are all taken at $t=0.02$. In-line with the speed-up the maximum fixation time is found at $\Da\approx 10^{5}$ for the noise flow, see Fig.~\ref{fig:fig1}. The data reveals an interesting effect; adding Brownian noise to the structured flow can accelerate fixation at $\Da\approx 10^0-10^3$, but leads to a slowdown of fixation at higher Damk\"ohler numbers, $\Da\approx10^4-10^6$.

\begin{figure}[t!!!!!]
\centering
  \includegraphics[scale = 0.375]{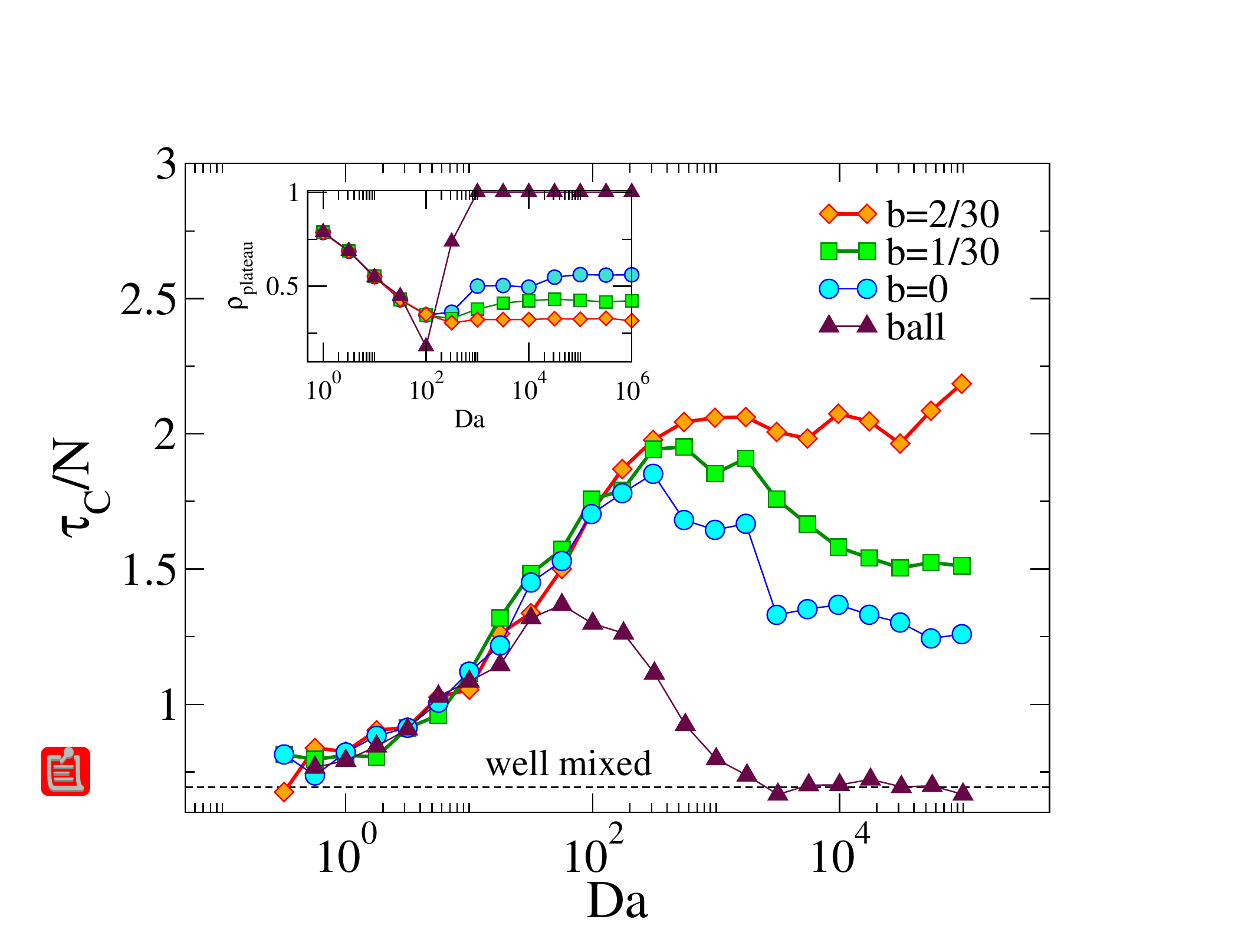}
  \caption{(Color on-line). Role of the initial network. Main panel: Average fixation time for initial networks obtained from a square lattice by random displacements of magnitude $b$. Inset: Order parameter $\rho$ before fixation. We also show data from simulations in which all particles are initially placed in a ball of diameter $\delta$.}
  \label{fig:fig4}
  \end{figure}

{\em Role of the initial network.} Focusing on the structured flow we have tested different initial networks, displacing the nodes of a regular grid by random amounts in $[-b,b]$. For $\Da\lesssim 10^2$ the flow is sufficiently fast to wash out the initial condition before the population reaches an absorbing state; as seen in Fig.~\ref{fig:fig4} the fixation time is not affected by the random displacements. For $\Da\gtrsim 10^2$ the initial network. Adding disorder to the initial grid increases the fixation time. A similar effect can be obtained by starting the flow from a regular grid, but initiating the evolutionary dynamics only after some transient time. This highlights the role of the flow in generating a modular network. Moderate flow can slow down fixation if the evolutionary dynamics starts from an ordered grid. This slowdown is due to the disorder generated by the flow. If the initial graph already contains sufficient disorder the flow does not create any additional modularity. It mainly stirs the nodes resulting in a monotonic reduction of fixation time with increased flow.  An interesting numerical experiment can be performed by initially placing all particles inside a ball of diameter $\delta$. This describes the situation of a densely populated colony, and constitutes a well mixed system in the no-flow limit. For very fast flow the system will effectively be well mixed as well. In either of the two limits the fixation time is $\tau_C=N\ln 2$. As seen in Fig.~\ref{fig:fig4} the interpolating behaviour is again non-monotonic, with significant slowdown of fixation at intermediate Damk\"ohler numbers.

{\em Conclusions.} We have studied the effects of a chaotic flow on a simple birth-death process in a finite population. The flow generates a dynamic interaction network, and genetic drift eventually leads to fixation of one species. If the initial particle positions form an ordered structure a slowdown of fixation occurs at intermediate Damk\"ohler numbers. Species can coexist the longest when the speed of network turnover is comparable to that of the macroscopic evolutionary process. This balance of time scales is in-line with proposed explanations of biodiversity \cite{dovidio, kudela}. We show that the effect in our model is due to heterogeneity in the interaction network, generated by the flow.  We have also tested a scenario in which an initially dense population is placed into a flowing medium (Fig. \ref{fig:fig4}), and another in which a small group of mutants with a fitness advantage invades a resident population \cite{gm}. Results again indicate that species coexist the longest at intermediate Damk\"ohler numbers. This suggests that the slowdown effect is robust, and that it may persist in less stylised models of marine and atmospheric ecologies with multiple time scales.


\begin{acknowledgments}
TG would like to thank the Group of Nonlinear Physics, University of Santiago de Compostela for hospitality. VPM acknowledges support by Ministerio de Econom\'{\i}a y Competitividad and Xunta de Galicia (MAT2015-71119-R, GPC2015/014), and contributions by the COST Action MP1305.
\end{acknowledgments}


\begin{thebibliography}{10}
\bibitem{gause} G. F. Gause, J. Exp. Biol. {\bf 9}, 389 (1932)
\bibitem{hardin} G. Hardin, Science {\bf 131}, 1292 (1960)
\bibitem{paradox}  G. E. Hutchinson, Am. Nat. {\bf 95}, 137 (1961).
\bibitem{karolyi2000} G. K\'arolyi, \'A. P\'entek, I. Scheuring, T. T\'el, and Z. Toroczkai, Proc. Nat. Acad. Sci. USA {\bf 97}, 13661 (2000).
\bibitem{young} W. Young, A. Roberts and G. Stuhne, Nature {\bf 412,} 328 (2001).
 \bibitem{neufeld2} Z. Neufeld, P. Haynes, and T. T\'el, Chaos {\bf 12}, 426 (2002).
 \bibitem{neufeld} G. K\'arolyi, Z. Neufeld, and I. Scheuring, J. Theor. Biol. {\bf 236}, 12 (2005).
\bibitem{groselj} D. Gro\v{s}elj, F. Jenko, and E. Frey, Phys. Rev. E {\bf 91}, 033009 (2015).
\bibitem{dovidio} F. d'Ovidio, S. De Monte, S. Alvain, Y. Dandonneau and M. Levy, Proc. Natl. Acad. Sci. USA {\bf 107}, 18366 (2010).
\bibitem{kudela} R. M. Kudela, Proc. Nat. Acad. Sci. USA {\bf 107}, 18235 (2010).
\bibitem{pasquero} C. Pasquero, Geophys. Res. Lett. {\bf 32}, L17603 (2005).
\bibitem{pigolotti2012} S. Pigolotti, R. Benzi, M.H. Jensen, and D.R. Nelson, Phys. Rev. Lett. {\bf 108}, 128102 (2012).
\bibitem{benzi2012} R. Benzi, M.H. Jensen, D.R. Nelson, P. Perlekar, S. Pigolotti, and F. Toschi, Eur. Phys. J. Special Topics {\bf 204}, 57 (2012).
\bibitem{perlekar2013} P. Perlekar, R. Benzi, D. R. Nelson, F. Toschi, and J. Turb. {\bf 14}, 161 (2013).
\bibitem{pigolotti2013} S. Pigolotti, R. Benzi, P. Perlekar, M.H. Jensen, F. Toschi, and D. R. Nelson, Theor. Pop. Biol. {\bf 84}, 72 (2013)
\bibitem{pigolotti2014} S. Pigolotti and R. Benzi, Phys. Rev. Lett. {\bf 112}, 188102 (2014).
\bibitem{abraham} E. R. Abraham, Nature {\bf 391}, 577 (1998).
\bibitem{ewens} W. J. Ewens, {\em Mathematical Population Genetics} (Springer, New York, 2004).
\bibitem{nowak} M. A. Nowak, {\em Evolutionary Dynamics} (Belknap Press, Harvard, 2006)

\bibitem{liggett} T. M. Liggett, {\em Interacting Particle Systems}, (Springer Verlag, New York, 1985).


\bibitem{fortunato} C. Castellano, S. Fortunato, and V. Loreto, Rev. Mod. Phys. {\bf 81}, 591 (2009).
\bibitem{blythe1} R. A. Blythe, A. J. McKane, J. .Stat. Mech.: Theo. and Exp. {\bf 2007}, P07018 (2007).
\bibitem{gross} T. Gross, H. Sayama (Eds), {\em Adaptive Networks: Theory, Models and Applications}, (Springer, Heidelberg, 2009).
\bibitem{fragmentation} F. V\'azquez, V. M. Egu\'iluz, and M. San~Miguel, Phys. Rev. Lett. {\bf 100}, 108702 (2008).




\bibitem{bennaim} E. Ben-Naim, L. Frachebourg, and P.L. Krapivsky, Phys. Rev. E {\bf 53}, 3078 (1996)
\bibitem{sood} V. Sood, T. Antal, and S. Redner, Phys. Rev. E {\bf 77}, 041121 (2008)





\bibitem{shear1} R. T. Pierrehumbert, Chaos, Solitons Fractals {\bf 4}, 1091 (1994).
\bibitem{shear2} V. P\'erez-Mun\~uzuri and G. Fern\'andez-Garc\'{\i}a, Phys. Rev. E {\bf 75}, 046209 (2007).
\bibitem{shear3} Z. Neufeld, E. Hern\'andez-Garc\'{\i}a, {\em Chemical and Biological Processes in Fluid Flows} (Imperial College Press, London, 2010).
\bibitem{remark1} The mean degree and clustering coefficient remain constant up to $t\approx 0.01$, but then vary. From $t\approx 5$ no further changes are observed.
\bibitem{gm} T. Galla and V. P\'erez-Mun\~uzuri (unpublished).





\end{thebibliography}
\end{document}